\documentclass[twocolumn,prb,showpacs,superscriptaddress,groupedaddress]{revtex4}
\usepackage{epsfig,color}
\usepackage{amssymb}
\usepackage{amsmath}
\usepackage{dcolumn}
\usepackage{graphicx}

\def\beq{\begin{equation}}
\def\eeq{\end{equation}}
\def\bea{\begin{eqnarray}}
\def\eea{\end{eqnarray}}

\def\ra{\rangle}
\def\la{\langle}

\def\k{\mathbf{k}}

\newcommand{\sgn}[1] {\mathrm{sgn}\left({#1}\right)}

\newcommand{\ii}{{\mathrm{i}}}

\newcommand{\iM}{\mathcal{I}}
\newcommand{\jM}{\mathcal{J}}

\newcommand{\imp}{\mathrm{imp}}

\begin{document}

\title{Unexpected impact of magnetic disorder on multiband superconductivity}
\author{M.M.~Korshunov}
\email{mkor@iph.krasn.ru}
\affiliation{Kirensky Institute of Physics, Siberian Branch of Russian Academy of Sciences, 660036 Krasnoyarsk, Russia}
\author{D.V.~Efremov}
\email{d.efremov@ifw-dresden.de}
\affiliation{Leibniz-Institut f\"{u}r Festk\"{o}rper- und Werkstoffforschung, D-01069 Dresden, Germany}
\author{A.A.~Golubov}
\affiliation{Faculty of Science and Technology and MESA+ Institute of Nanotechnology, University of Twente, 7500 AE Enschede, The Netherlands}
\affiliation{Moscow Institute of Physics and Technology, Dolgoprudnyi, Russia}
\author{O.V.~Dolgov}
\affiliation{Max-Planck-Institut f\"{u}r Festk\"{o}rperforschung, D-70569 Stuttgart, Germany}
\affiliation{Lebedev Physical Institute RAN, Moscow, Russia}

\date{\today}

\begin{abstract}
We analyze how the magnetic disorder affects the properties of the two-band $s_\pm$ and $s_{++}$ models, which are subject of hot discussions regarding iron-based superconductors and other multiband systems like MgB$_2$. We show that there are several cases when the transition temperature $T_c$ is not fully suppressed by magnetic impurities in contrast to the Abrikosov-Gor'kov theory, but a saturation of $T_c$ takes place in the regime of strong disorder. These cases are: (1) the purely interband impurity scattering, (2) the unitary scattering limit. We show that in the former case the $s_\pm$ gap is preserved, while the $s_{++}$ state transforms into the $s_\pm$ state with increasing magnetic disorder. For the case (2), the gap structure remains intact.
\end{abstract}

\maketitle

\textit{Introduction.} Since the discovery of Fe-based superconductors (FeBS) in 2008~\cite{kamihara}, the main question remains open: what is the driving mechanism for superconductivity in this class of materials? Excluding the cases of extreme hole and electron dopings, the Fermi surface (FS) consisting of two or three hole pockets around the $\Gamma=(0,0)$ point and two electron pockets around the $M=(\pi,\pi)$ point in the 2-Fe Brillouin zone naturally leads to the enhanced antiferromagnetic fluctuations. They lead to the $s$-wave-like order parameter that change sign between electron and hole pockets, the so-called $s_\pm$ state~\cite{reviews}. On the other hand, bands near the Fermi level have mixed orbital content and orbital fluctuations enhanced by the electron-phonon coupling may lead to the sign-preserving $s$-wave gap, the $s_{++}$ state~\cite{kontani,bang}. Most experimental data including observation of the spin-resonance peak in inelastic neutron scattering, quasiparticle interference in tunneling experiments, and NMR spin-lattice relaxation rate are in favor of the $s_\pm$ scenario, although gap anisotropy varies from one material to the other~\cite{reviews}. Therefore, an experimental probe that can uniquely pinpoint the gap structure is of high demand.

It was suggested some time ago that the scattering on impurities may disentangle sign-changing and sign-preserving gaps~\cite{kontaniimp,kontaniimspm,efremov,Efremov2013,Dolgov2013}. Usually, nonmagnetic impurity scattering between the bands with different signs of the gaps leads to suppression of the critical temperature $T_c$ similar to magnetic impurity scattering in a single-band BCS superconductor~\cite{golubov97}. Then $T_c$ is determined from the Abrikosov-Gor'kov (AG) expression $\ln T_{c0}/T_{c} = \Psi(1/2+ \Gamma/2\pi T_c) - \Psi(1/2)$, where $\Psi(x)$ is the digamma function, $\Gamma$ is the impurity scattering rate, and $T_{c0}$ is the critical temperature in the absence of impurities~\cite{AG}. However, it was recently shown that in the multiband superconductors the behavior may be more complicated~\cite{golubov97,Ummarino2007,efremov}. In particular, $T_c$ is almost constant for varying the amount of nonmagnetic disorder in i) unitary limit of the uniform intra- vs. interband scattering potentials~\cite{dolgovkulic} and ii) $s_\pm \to s_{++}$ transition for the sizeable intraband attraction in the two-band $s_\pm$ model in the strong-coupling $T$-matrix approximation~\cite{efremov}. Qualitatively these results were confirmed via the numerical solutions of the Bogoliubov-de Gennes equations~\cite{Yao2012,Chen2013}. Several experiments show that the $T_c$ suppression is much weaker than expected in the framework of the AG theory for both non-magnetic~\cite{Cheng2010,li,nakajima,tropeano,Kim2014,Prozorov2014} and magnetic disorder~\cite{Cheng2010,Tarantini2010,Tan2011,Grinenko2011,Li2012}.
Overall, there is no clear answer to whether the sign-changing order parameter symmetry is prevailing in FeBS. In such situation, additional information can be gained from studies of the magnetic impurities.


Here we study two-band models for the isotropic $s_\pm$ and $s_{++}$ superconductors with the magnetic impurities in the self-consistent $T$-matrix approximation~\cite{allen}. We argue that $T_c$ and order parameter dependencies on the concentration of magnetic impurities have some peculiarities, which may help to identify the order parameter in the clean case. The common wisdom is that the magnetic impurities should suppress superconductivity. We show that while in general there is a suppression of superconducting state with increasing concentration of magnetic disorder, there are several regimes with either the absence of the $T_c$ suppression or the drastic reduction of this effect.
But even if $T_c$ is completely suppressed, its behavior may differ from the AG theory for the single-band superconductors.
Other prototypical examples of multiband systems where our theory is applicable include MgB$_2$ and the approximate treatment of the $d$-wave superconductors like cuprates where the parts of the Fermi surface with the different signs of the gap to some extent can be considered as different bands.

\section{Method} 

We employ the Eliashberg approach for multiband superconductors~\cite{allen} and calculate the $\xi$-integrated Green's functions $\hat{\mathbf{g}}(\omega_n) = \int d \xi \hat{\mathbf{G}}(\k, \omega_n) =
\left(
\begin{array}{cc}
\hat{g}_{an} & 0 \\
0 & \hat{g}_{bn}
\end{array}
\right)$,
%
%
%
%
where $\hat{g}_{\alpha n} = g_{0\alpha n} \hat{\tau}_{0}\otimes \hat{\sigma}_{0} + g_{2\alpha n} \hat{\tau}_{2}\otimes \hat{\sigma}_{2}$, indices $a$ and $b$ correspond to two distinct bands, index $\alpha = a,b$ denote the band space, Pauli matrices define Nambu ($\hat{\tau}_{i}$) and spin ($\hat{\sigma}_{i}$) spaces, $\hat{\mathbf{G}}(\k,\omega_n) = \left[\hat{\mathbf{G}}_0^{-1}(\k,\omega_n) - \hat{\mathbf{\Sigma}}(\omega_n)\right]^{-1}$ is the matrix Green's function for a quasiparticle with momentum $\k$ and the Matsubara frequency $\omega_n = (2 n + 1) \pi T$ defined in the band space and in the combined Nambu and spin spaces, $\hat{G}_0^{\alpha \beta}({\mathbf{k}},\omega_n)=\left[ \ii \omega_n \hat{\tau}_{0}\otimes \hat{\sigma}_{0}-\xi_{\alpha \k}\hat{\tau}_{3}\otimes \hat{\sigma}_{0}\right]^{-1} \delta_{\alpha \beta}$ is the bare Green's function,
$\hat{\mathbf{\Sigma}}(\omega _{n}) = \sum_{i=0}^{3} \Sigma_{\alpha \beta}^{(i)}(\omega_n)\hat{\tau}_i$ is the self-energy matrix, $\xi_{\alpha, \k} = v_{\alpha, F} (k-k_{\alpha, F})$ is the linearized dispersion, $g_{0\alpha n}$ and $g_{2\alpha n}$ are the normal and anomalous $\xi$-integrated Nambu Green's functions,
\beq
g_{0\alpha n}=-\frac{\ii \pi N_{\alpha} \tilde{\omega}_{\alpha n}}{\sqrt{\tilde{\omega}_{\alpha n}^{2}+\tilde{\phi}_{\alpha n}^{2}}}, \;\;\; g_{2\alpha n}=-\frac{\pi N_{\alpha} \tilde{\phi}_{\alpha n}}{\sqrt{\tilde{\omega}_{\alpha n}^{2}+\tilde{\phi}_{\alpha n}^{2}}},
\label{g}
\eeq
depending on the density of states per spin of the corresponding band at the Fermi level $N_{a,b}$ and on renormalized (by the self-energy) order parameter $\tilde{\phi}_{\alpha n}$ and frequency $\tilde{\omega}_{\alpha n}$,
\begin{eqnarray}
\ii \tilde\omega_{a n} &=& \ii \omega_n -  \Sigma_{0a}(\omega_n) - \Sigma_{0a}^{\imp}(\omega_n), \label{eq.omega.tilde} \\
\tilde\phi_{a n} &=& \Sigma_{2a}(\omega_n) + \Sigma_{2a}^{\imp}(\omega_n). \label{eq.Delta.tilde}
\end{eqnarray}
It is also convenient to introduce the renormalization factor $Z_{\alpha n} = \tilde{\omega}_{\alpha n} / \omega_n$ that enters the gap function $\Delta_{\alpha n} = \tilde{\phi}_{\alpha n} / Z_{\alpha n}$. The self-energy due to the spin fluctuation interaction is then given by
\bea
\Sigma_{0\alpha}(\omega_n) &=& T \sum\limits_{\omega_n',\beta} \lambda^{z}_{\alpha\beta} (n-n') \frac{g_{0\beta n}}{N_\beta}, \label{eq.DeltaN2} \\
\Sigma_{2\alpha}(\omega_n) &=& -T \sum\limits_{\omega_n',\beta} \lambda^{\phi}_{\alpha\beta}(n-n') \frac{g_{2\beta n}}{N_\beta},
\label{eq.DeltaN1}
\eea
The coupling functions $\lambda^{\phi,z}_{\alpha\beta}(n-n') = 2 \lambda^{\phi,z}_{\alpha\beta} \int^{\infty}_{0} d\Omega \Omega B(\Omega) / \left[(\omega_n-\omega_{n'})^{2} + \Omega^{2}\right]$ depend on the normalized bosonic spectral function
$B(\Omega)$ used in Refs.~\onlinecite{efremov,Efremov2013}.
While the matrix elements $\lambda^\phi_{\alpha \beta}$  can be positive (attractive) as well as negative (repulsive) due to the interplay between spin fluctuations and electron-phonon coupling~\cite{BS,parker}, the matrix elements $\lambda^z_{\alpha \beta}$ are always positive. For simplicity we set $\lambda^z_{\alpha \beta}=|\lambda^\phi_{\alpha \beta}|\equiv|\lambda_{\alpha \beta}|$ and neglect possible anisotropy in each order parameter $\tilde\phi_{\alpha n}$. Latter effects can lead to changes in the response of the two-band $s_\pm$ system to disorder and have been examined, e.g. in Ref.~\onlinecite{Mishra}.

We use the $T$-matrix approximation to calculate the average impurity
self-energy $\hat{\mathbf{\Sigma}}^{\imp}$:
\begin{equation}
\hat{\mathbf{\Sigma}}^{\imp}(\omega_n) = n_{\imp} \hat{\mathbf{U}} + \hat{\mathbf{U}} \hat{\mathbf{g}}(\omega_n) \hat{\mathbf{\Sigma}}^{\imp}(\omega_n),
\label{eq.tmatrix}
\end{equation}
where $n_{\imp}$ is the impurity concentration. Impurity potential for the non-correlated impurities can be written as $\hat{\mathbf{U}}=\mathbf{V} \otimes \hat{S}$, where
$ \hat{S} = \mathrm{diag}\left[\vec{\hat{\sigma}} \cdot \vec{S}, -(\vec{\hat{\sigma}} \cdot \vec{S})^{T}\right]$
%
%
is the $4 \times 4$ matrix with $(...)^{T}$ being the matrix transpose and $\vec{S} = \left( S_x, S_y, S_z \right)$ being the spin vector~\cite{ambeg}. The vector $\vec{\hat{\sigma}}$ is composed of $\tau$ matrices, $\vec{\hat{\sigma}} = \left( \hat{\tau}_1, \hat{\tau}_2, \hat{\tau}_3 \right)$. The potential strength is determined by $(\mathbf{V})_{\alpha \beta} = V_{\mathbf{R}_i = 0}^{\alpha \beta}$.
For simplicity intraband and interband parts of the potential are set equal to $\iM$ and $\jM$, respectively, such that $(\mathbf{V})_{\alpha \beta} = (\iM-\jM) \delta_{\alpha \beta} + \jM$.
Components of the impurity potential matrix $\hat{\mathbf{U}}$ is then $\hat{U}_{aa,bb} = \iM \hat{S}$ and $\hat{U}_{ab,ba} = \jM \hat{S}$.
Coupled $T$-matrix equations for $aa$ and $ba$ components of the self-energy become
\bea
\hat{\Sigma}_{aa}^{\imp} &=& n_{\imp} \hat{U}_{aa} + \hat{U}_{aa} \hat{g}_a \hat{\Sigma}_{aa}^{\imp} + \hat{U}_{ab} \hat{g}_b \hat{\Sigma}_{ba}^{\imp},
\label{eq.Sigma_aa} \\
\hat{\Sigma}_{ba}^{\imp} &=& n_{\imp} \hat{U}_{ba} + \hat{U}_{ba} \hat{g}_a \hat{\Sigma}_{aa}^{\imp} + \hat{U}_{bb} \hat{g}_b \hat{\Sigma}_{ba}^{\imp}.
\label{eq.Sigma_ba}
\eea
Renormalizations of frequencies and gaps come from $\Sigma^{\imp}_{0a} = \frac{1}{4} \mathrm{Tr}\left[\hat{\Sigma}_{aa}^{\imp} \cdot \left( \hat{\tau}_0 \otimes \hat{\sigma}_0 \right) \right]$ and $\Sigma^{\imp}_{2a} = \frac{1}{4} \mathrm{Tr}\left[\hat{\Sigma}_{aa}^{\imp} \cdot \left( \hat{\tau}_2 \otimes \hat{\sigma}_2 \right) \right]$.


\section{Results} 

Following results were obtained by solving self-consistently frequency and gap equations~(\ref{eq.omega.tilde}) and (\ref{eq.Delta.tilde}) with the impurity self-energy from the solution of Eqs.~(\ref{eq.Sigma_aa}), (\ref{eq.Sigma_ba}) for both finite temperature and at $T_c$. Expressions for $\Sigma^{\imp}_{0\alpha}$ and $\Sigma^{\imp}_{2\alpha}$ are proportional to the effective impurity scattering rate $\Gamma_{a,b}$ and as in Ref.~\onlinecite{efremov} contain the generalized cross-section parameter $\sigma$ that helps to control the approximation for the impurity strength ranging from Born (weak scattering, $\pi \jM N_{a,b} \ll 1$) to the unitary (strong scattering, $\pi \jM N_{a,b} \gg 1$) limits,
\bea
\Gamma_{a,b} &=& \frac{2 n_{\imp} \sigma}{\pi N_{a,b}} \to \left\{
    \begin{array}{l}
    2 \pi \jM^2 s^2 n_{\imp} N_{b,a}, \text{Born}\\
    \frac{2 n_{\imp}}{\pi  N_{a,b}}, \text{unitary}
    \end{array}
  \right.
\\
\sigma &=& \frac{\pi^2 \jM^2 s^2 N_a N_b}{1 + \pi^2 \jM^2 s^2 N_a N_b}
\to \left\{
    \begin{array}{l}
    0, \text{Born}\\
    1, \text{unitary}
    \end{array}
  \right.
\eea
Note that $\Gamma_{\alpha}$ here is twice as large as defined in Ref.~\onlinecite{efremov}. We assume that spins are not polarized
and $s^2 = \la S^2 \ra = S(S+1)$.
Also, we introduce the parameter $\eta$ to control the ratio of intra- and interband scattering potentials, $\iM = \jM \eta$.

\begin{figure}[t]
\centering
\includegraphics[width=1\columnwidth]{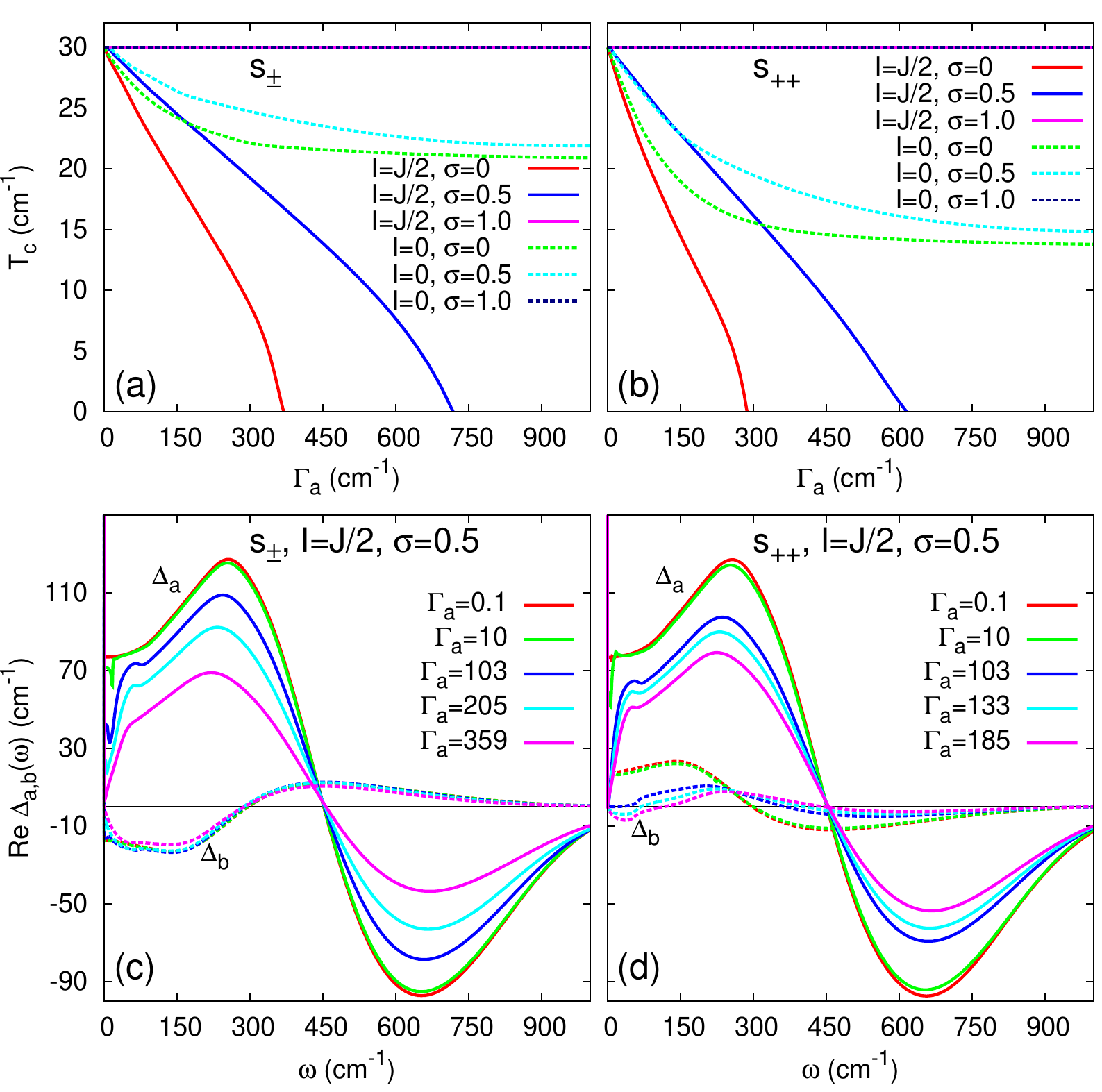}
\caption{(color online). $T_{c}$ dependence on the scattering rate $\Gamma_a$ (a,b) and frequency dependence of gaps $\mathrm{Re}\Delta_\alpha(\omega)$ (c,d) for various values of $\Gamma_a$ for the $s_\pm$ (a,c) and the $s_{++}$ (b,d) superconductors. $N_{b}/N_{a}=2$ and coupling constants are $(\lambda_{aa},\lambda_{ab},\lambda_{ba},\lambda_{bb}) = (3,-0.2,-0.1,0.5)$ so that $\la \lambda \ra < 0$ for the $s_\pm$ state and $(3,0.2,0.1,0.5)$ for the $s_{++}$ state.}
\label{fig:spmsppTc}
\end{figure}
\begin{figure}[t]
\centering
\includegraphics[width=1\columnwidth]{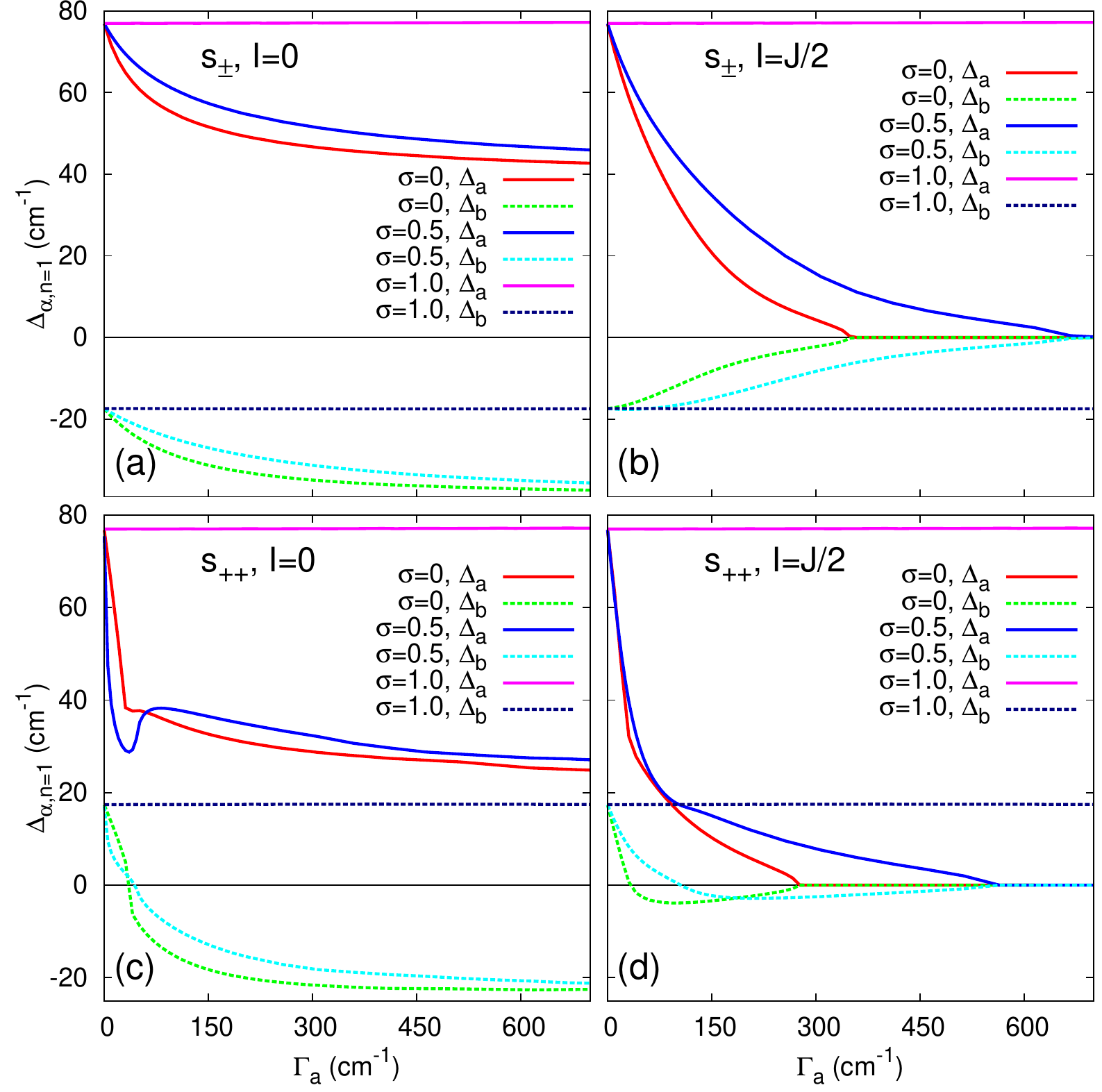}
\caption{(color online). Matsubara gap $\Delta_{\alpha n=1}$ dependence on the scattering rate $\Gamma_a$ for the $s_\pm$ (a,b) and the $s_{++}$ (c,d) superconductors with only interband scattering, $\iM=0$ (a,c), and with $\iM=\jM/2$ (b,d). Parameters are the same as in Fig.~\ref{fig:spmsppTc}.}
\label{fig:spmsppDelta}
\end{figure}

In Fig.~\ref{fig:spmsppTc}(a,b) and~\ref{fig:spmsppDelta} we plot $T_c$ and the gap function $\Delta_{\alpha n}$ for the first Matsubara frequency $\omega_{n=1} = 3 \pi T$
vs. $\Gamma_a$ for a set of $\sigma$'s for both $s_\pm$ and $s_{++}$ superconductors. Real part of the analytical continuation of $\Delta_{\alpha n}$ to real frequencies, the gap function $\mathrm{Re}\Delta_\alpha(\omega)$, is shown in Fig.~\ref{fig:spmsppTc}(c,d). First, we discuss the $s_\pm$ state. $T_c$ becomes insensitive to impurities for the pure interband scattering, $\iM = 0$. This partially confirms qualitative arguments that $s_\pm$ state with magnetic disorder behave like the $s_{++}$ state with non-magnetic impurities~\cite{golubov97} and agrees with the quantitative theoretical calculations in the Born limit~\cite{Li2009}. For the finite $\iM$, intraband scattering on magnetic disorder average gaps to zero thus suppressing $T_c$. On the other hand, in the unitary limit ($\sigma=1$) at $T \to T_c$ we have $\tilde\omega_{a n} = \omega_n + \ii \Sigma_{0a}(\omega_n) + \frac{\Gamma_a}{2} \sgn{\omega_n}$ and $\tilde\phi_{a n} = \Sigma_{2a}(\omega_n) + \frac{\Gamma_a}{2} \frac{\tilde\phi_{a n}}{\left|\tilde\omega_{a n}\right|}$ for any value of $\eta$ including the case of intraband-only impurities,
$1/\eta = 0$. This form is the same as for non-magnetic impurities and thus analogously to the Anderson theorem there is no impurity contribution to the $T_c$ equation. The only exception here is the special case of uniform impurities, $\eta = 1$, when $\tilde\omega_{a n} = \omega_n + \ii \Sigma_{0a}(\omega_n) + \frac{n_{\imp}}{\pi  \left(N_a+N_b\right)} \sgn{\omega_n}$ and $\tilde\phi_{a n} = \Sigma_{2a}(\omega_n) + \frac{n_{\imp}}{\pi \left(N_a+N_b\right)^2} \left(N_a \frac{\tilde\phi_{a n}}{\left|\tilde\omega_{a n}\right|} + N_b \frac{\tilde\phi_{b n}}{\left|\tilde\omega_{b n}\right|} \right)$. Both gaps are mixed in equation for $\tilde\phi_{a n}$, thus they tend to zero with increasing amount of disorder. That's also true away from the unitary limit and that's why there is a special case of uniform potential of the impurity scattering, $\iM = \jM$, when the strongest $T_c$ suppression occurs. For the initially unequal gaps, $|\Delta_a| \neq |\Delta_b|$, there is an initial decrease of $T_c$ for small $\Gamma_a$ until the renormalized gaps become equal and then $T_c$ saturate since the analog of Anderson theorem achieved.
\begin{figure}[t]
\centering
\includegraphics[width=0.9\columnwidth]{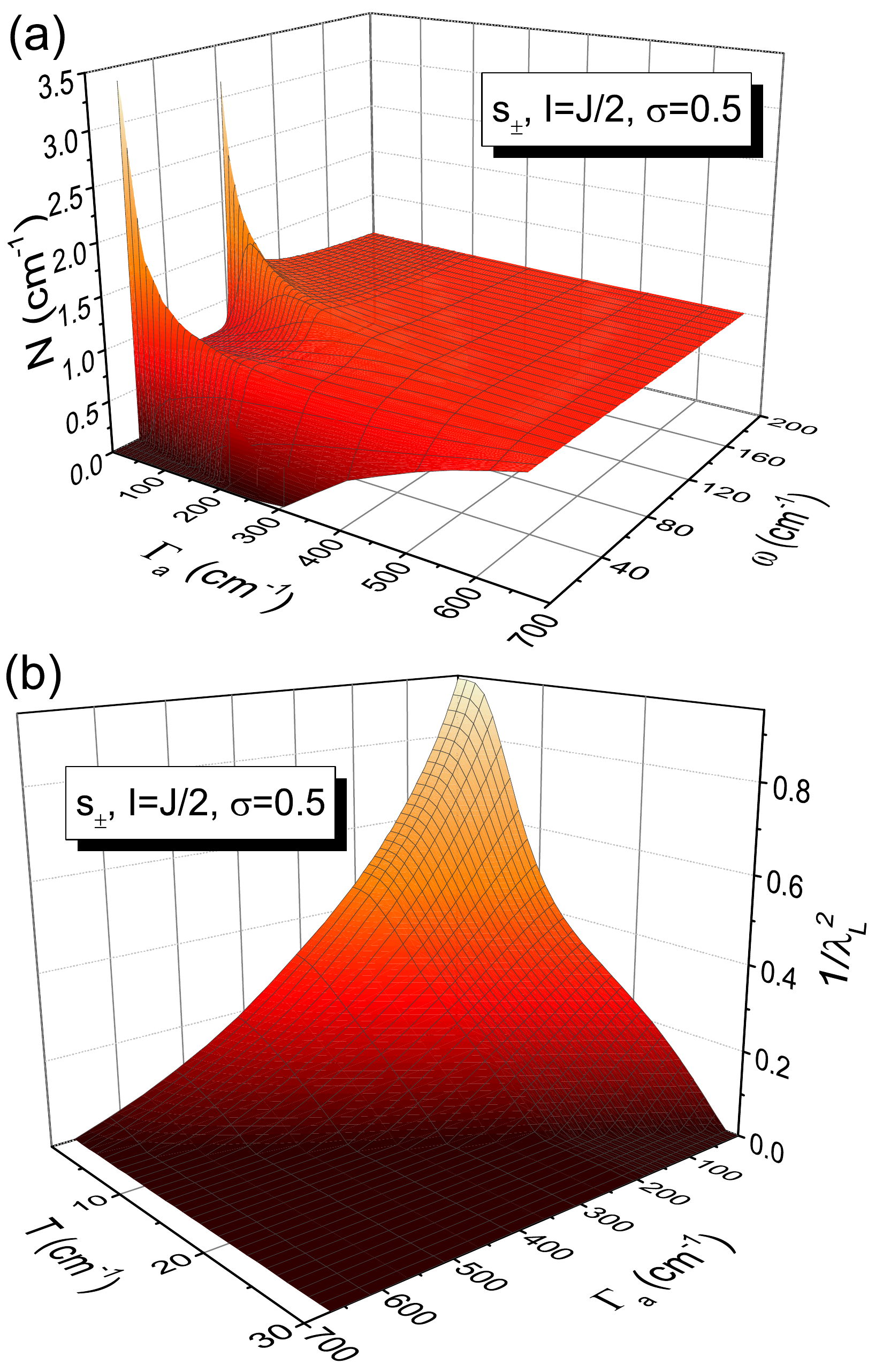}
\caption{(color online) Density of states $N$ as a function of frequency $\omega$ and interband magnetic impurities scattering rate $\Gamma_a$ (a) and inverse squared penetration depth $1/\lambda_{L}^2$ vs. $\Gamma_{a}$ and $T$ (b) for the $s_{\pm}$ superconductor with $\iM=\jM/2$, $\sigma=0.5$, and parameters as in Fig.~\ref{fig:spmsppTc}.}
\label{fig:dosspm}
\end{figure}
\begin{figure}[t]
\centering
\includegraphics[width=0.9\columnwidth]{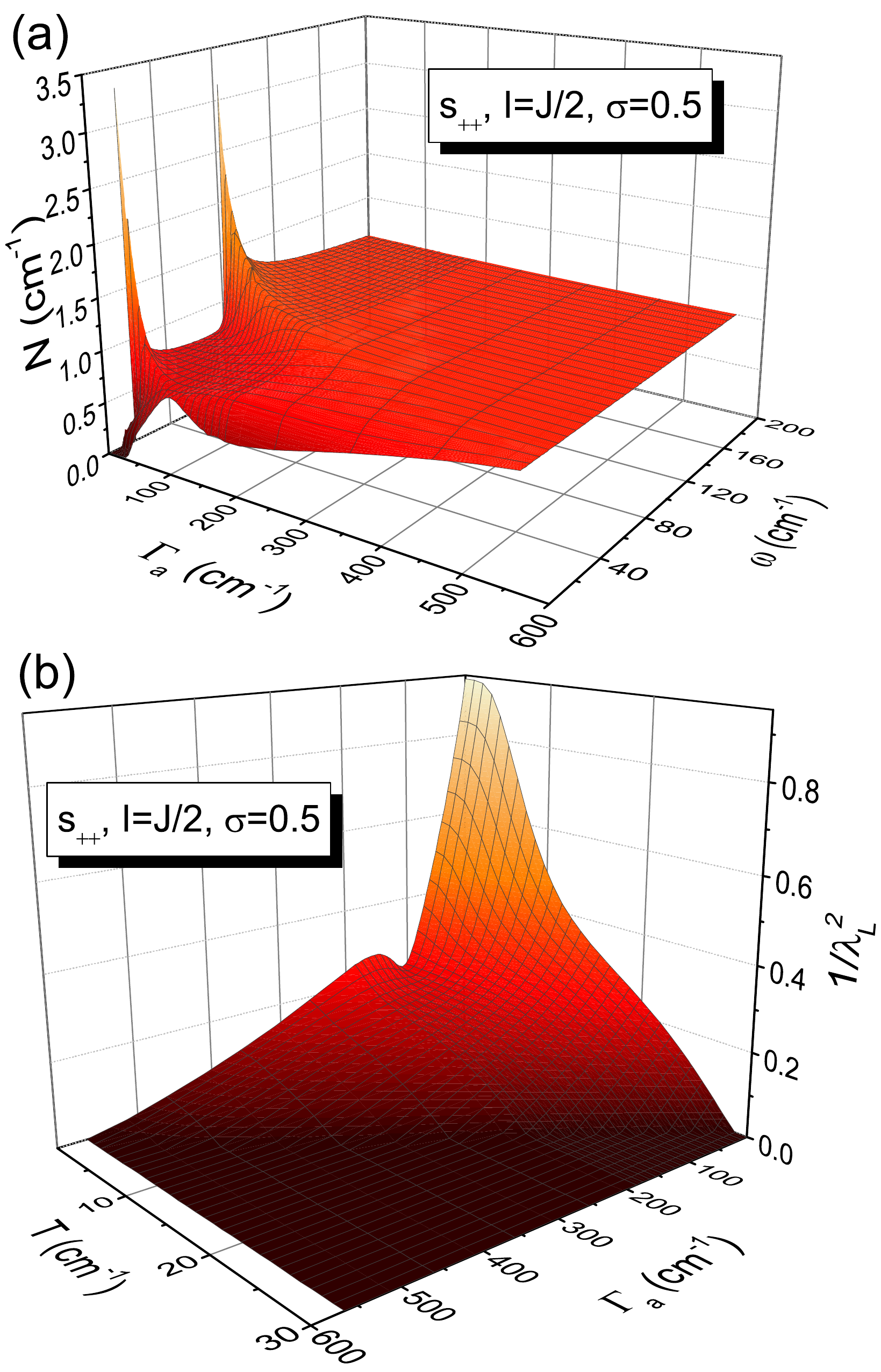}
\caption{(color online) Density of states $N$ as a function of frequency $\omega$ and $\Gamma_a$ (a) and inverse squared penetration depth $1/\lambda_{L}^2$ vs. $\Gamma_{a}$ and $T$ (b) for the $s_{++}$ superconductor with $\iM=\jM/2$, $\sigma=0.5$, and parameters as in Fig.~\ref{fig:spmsppTc}. Note the $s_{++}$ to $s_\pm$ transition at $\Gamma_a \sim 100$~cm$^{-1}$ and the gapless region right after that.}
\label{fig:dosspp}
\end{figure}

In general, multiband $s_{++}$ state should always be fragile against paramagnetic disorder since magnetic scattering between bands of the same sign effectively equivalent to the pairbreaking scattering within the single (quasi)isotropic band. Surprisingly, we find a regime with the saturation of $T_c$ for the finite amount of disorder right after the initial AG downfall, see Fig.~\ref{fig:spmsppTc}(b). The saturation of $T_c$ is observed for the interband-only impurities; presence of the intraband magnetic disorder finally suppress $T_c$ to zero. But depending on the ``strength'' of scattering $\sigma$, decrease of $T_c$ may be quite slow compared to the AG law.

To understand the origin of the $T_c$ saturation we analyzed the gap function dependence on the scattering rate $\Gamma_a$, see Fig.~\ref{fig:spmsppDelta}. For the $s_{++}$ state after the certain value of the scattering rate the smaller gap, $\Delta_b$, becomes negative. What we see is the $s_{++} \to s_\pm$ transition. As soon as system becomes effectively $s_\pm$, the scattering on magnetic impurities cancels out in the $T_c$ equation similar to the Anderson theorem and $T_c$ saturates. Before saturation, the initial AG downfall takes place. The transition is also seen in the gap function on real frequencies, Fig.~\ref{fig:spmsppTc}(d).

Similar to the $s_\pm \to s_{++}$ transition for the non-magnetic disorder, there is a simple physical argument behind the $s_{++} \to s_\pm$ transition here. Namely, with increasing interband magnetic disorder, the gap functions on the different Fermi surfaces tend to the same value and if one of the gaps is smaller than another, it cross zero and change sing. A similar effect has been mentioned in Refs.~\cite{scharnberg,golubov97} for a two-band systems with $s_{++}$ symmetry in the Born limit. Note that here we do not consider possible time-reversal symmetry broken $s_\pm + \ii s_{++}$ state that may be energetically favorable below $T_c$ in cases when translational symmetry is broken~\cite{Stanev2012}.

Since one of the gaps changes sign it necessary goes through zero. That corresponds to the gapless superconductivity. Therefore, the transition should manifest itself in the density of states measurable by tunneling and ARPES $N(\omega) = -\sum_{\alpha} \mathrm{Im} g_{0\alpha}(\omega)/\pi$, where $g_{0\alpha}(\omega)$ is the retarded Green's function, and in the temperature dependence of the London penetration depth $\lambda_{L}$, $\frac{1}{\lambda_{L}^2} = \sum\limits_{\alpha} \frac{\omega_{P\alpha}^2}{c^2} T \sum\limits_{n} \frac{ g_{2\alpha n}^2}{\pi N_{\alpha}^2 {\sqrt{\tilde{\omega}_{\alpha n}^{2}+\tilde{\phi}_{\alpha n}^{2}}} }$, where $\omega_{P\alpha}/c$ is the ratio of the plasma frequency to the sound velocity that we set to unity for simplicity.
In Fig.~\ref{fig:dosspm} and~\ref{fig:dosspp} we show $N(\omega)$ and $1/(\omega_{p}\lambda_{L})^{2}$ for the case of $\iM = \jM/2$ and $\sigma = 0.5$ for $s_\pm$ and $s_{++}$ superconductors. In the former case, Fig.~\ref{fig:dosspm} reflects the expected situation of the gradually decreasing gaps. The gapless superconductivity with a finite residual $N(\omega=0)$ appears for $\Gamma_a > 300$~cm$^{-1}$ when $\mathrm{Re}\Delta_{\alpha}(\omega=0)$ vanishes, see~\ref{fig:spmsppTc}(c). As for the $s_{++}$ case in Fig.~\ref{fig:dosspp}, with increasing impurity scattering rate $\Gamma_a$, the smaller gap vanishes leading to a finite residual $N(\omega=0)$. Then the gap reopens and $\Delta_{b n} \neq 0$ until $T_c$ reaches zero for $\Gamma_a \sim 600$~cm$^{-1}$, but the superconductivity remains gapless with finite $N(0)$ due to the $\mathrm{Re}\Delta_{\alpha}(\omega=0) \to 0$, see Fig.~\ref{fig:spmsppTc}(d). Penetration depth in the clean limit shows the activated behavior controlled by the smaller gap. For the $s_{++}$ superconductor it goes to the $T^{2}$ behavior in the gapless regime showing a pronounced dip in Fig.~\ref{fig:dosspp} around $\Gamma_a=100$~cm$^{-1}$ and crosses over to a new activated behavior in the $s_{+-}$ state after the transition.

\section{Conclusions} 

We have shown that contrary to the common wisdom in two-band models few exceptional cases exist with the saturation of $T_c$ for the finite amount of magnetic disorder. The particular case is the $s_\pm$ state in the unitary limit or with the purely interband impurity scattering potential. The latter satisfies qualitative assessment of direct relation between magnetic impurities in $s_\pm$ state and non-magnetic impurities in isotropic $s$-wave state. We demonstrate that $s_{++}$ superconductivity may be robust against magnetic impurities with the purely interband scattering due to the transition to the $s_\pm$ state. Since this transition goes through the gapless regime, there should be clear signatures in the thermodynamics of the system. Therefore, it may manifest itself in optical and tunneling experiments, as well as in a photoemission and thermal conductivity on FeBS and other multiband systems.

The authors are grateful to S.-L. Drechsler, P.J. Hirschfeld, K. Kikoin, and S.G. Ovchinnikov for useful discussions. We acknowledge partial support by the Dynasty Foundation and ICFPM (MMK), the Ministry of Education and Science of the Russian Federation (Grant No. 14Y26.31.0007), RFBR (Grants 12-02-31534 and 13-02-01395), President Grant for Government Support of the Leading Scientific Schools of the Russian Federation (NSh-2886.2014.2), DFG Priority Program 1458, and FP7 EU-Japan program IRON SEA.

\end{document}